\documentclass[
 reprint,
 aps,
 pra,
 amsmath,
 amssymb,
 superscriptaddress,
 longbibliography,
 showkeys,
 floatfix
]{revtex4-2}

\usepackage[T1]{fontenc}
\usepackage[utf8]{inputenc}
\usepackage{amsmath,amssymb,amsfonts,bm}
\usepackage{mathtools}
\usepackage{graphicx}
\usepackage{booktabs}
\usepackage{multirow}
\usepackage{xcolor}
\usepackage{hyperref}

\newcommand{\ket}[1]{\left|#1\right\rangle}

\hypersetup{
    colorlinks=true,
    linkcolor=blue,
    citecolor=blue,
    urlcolor=blue
}

\begin{document}

\title{Empirical Scaling Laws of Quantum Vision Transformers}

\author{Yuqing Li}
\email{yul658@pitt.edu}
\affiliation{Department of Computer Science, University of Pittsburgh, Pittsburgh, Pennsylvania 15260, USA}

\author{Fengyi Gao}
\email{feg38@pitt.edu}
\affiliation{Department of Information Science, University of Pittsburgh, Pittsburgh, Pennsylvania 15260, USA}

\author{Xin Jin}
\affiliation{Department of Computer Science, University of Pittsburgh, Pittsburgh, Pennsylvania 15260, USA}

\author{Xiaowei Jia}
\email{xiaowei.jia@rutgers.edu}
\affiliation{Department of Computer Science, Rutgers, The State University of New Jersey, New Brunswick, New Jersey 08901, USA}

\author{Qi Cheng}
\email{q.cheng@rutgers.edu}
\affiliation{Department of Computer Science, Rutgers, The State University of New Jersey, New Brunswick, New Jersey 08901, USA}

\author{Junyu Liu}
\email{junyuliu@pitt.edu}
\affiliation{Department of Computer Science, University of Pittsburgh, Pittsburgh, Pennsylvania 15260, USA}

\date{\today}

\begin{abstract}
Quantum machine learning (QML) is a promising application of quantum computing, yet when and how practical learning models can achieve quantum advantage remains poorly understood. Even a polynomial improvement in the resources needed to attain a given predictive performance would be significant. In this work, we investigate empirical scaling laws of quantum vision transformers (QViTs), which connect QML to the transformer architecture underlying modern large language models. We ask whether QViTs exhibit scaling behavior analogous to classical models and how this behavior changes when qubit number is included alongside dataset size and model parameters. The validation losses in our numerical image-classification experiments are well described by a joint shifted power-law model in qubit number, transformer depth, and training dataset size. Expressing this fit in terms of the classical encoder parameter budget identifies a regime in which increasing the number of qubits can compensate for fewer classical encoder parameters at comparable predicted performance. This finding is consistent with earlier observations of parameter efficiency in quantum learning models and suggests a direction for exploring possible resource savings in QML.
\end{abstract}


\maketitle

\section{Introduction}
Quantum machine learning (QML) is a promising application of quantum computing, with the potential to improve data representation and resource efficiency in machine learning. 
Despite substantial progress, the theory needed to identify useful quantum advantages in practical learning tasks remains incomplete. Rigorous advantages have been established for specific learning problems under computational hardness assumptions~\cite{Liu2021RigorousSpeedup}, while the role of training data makes the comparison between quantum and classical learning more subtle~\cite{Huang2021PowerData}. Quantum neural tangent kernels and wide-network analyses provide theoretical descriptions of variational-circuit training in specific regimes~\cite{Liu2022QNTK,Liu2023WideQNN}. Connecting this progress to predictive performance and resource scaling in practical architectures remains challenging. A central question is therefore where to seek quantum advantage and whether even a polynomial improvement in a relevant resource can be achieved at comparable predictive performance.

\begin{figure*}[t]
    \centering
    \includegraphics[width=\textwidth]{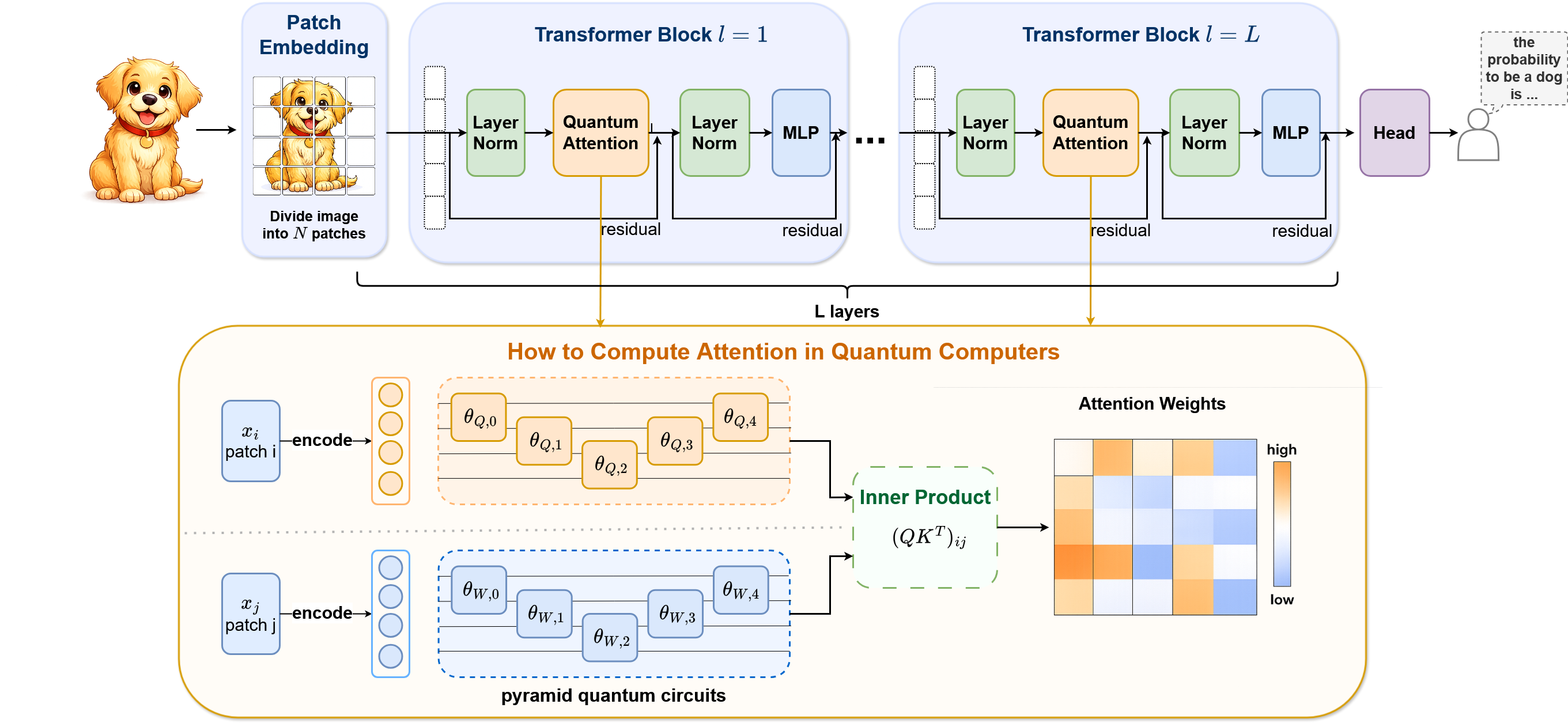}
    \caption{Structure of the QViT architecture. The top panel shows the full vision transformer pipeline from patch embedding to stacked transformer blocks and the final prediction head. The bottom panel illustrates the quantum attention module, in which encoded patch representations are first mapped to quantum states and then processed by parameterized quantum circuits composed of RBS gates to generate the query and key matrices, \(Q\) and \(K\).}
    \label{fig:QViT_overview}
\end{figure*}

Transformers provide a natural setting for studying 
this question. They underpin modern large language models (LLMs), and their scaling behavior has helped guide the development of large-scale AI systems~\cite{Kaplan2020ScalingLaws}. Quantum vision transformers (QViTs)~\cite{Cherrat2024quantumvision} bring quantum attention into this architecture, providing a direct connection between QML and the transformer models central to current LLM research. Complementary algorithmic work investigates quantum acceleration of large-scale model training under sparsity and dissipation assumptions~\cite{Liu2024LargeScaleQML}. Vision tasks offer a practical testbed for numerically studying quantum transformer models. Existing QViT studies include biomedical image classification~\cite{Boucher2025qsaBiomed} and high-energy physics~\cite{Cara2024qvitQuarkGluon,Unlu2024hybridQViT_HEP}; our focus is the scaling behavior that links these architectures to their resource requirements.

Classical neural scaling laws relate predictive performance to dataset size, parameter count, and training compute~\cite{Hestness2017Scaling,Zhai2022ScalingViT,Kaplan2020ScalingLaws}. A quantum model introduces an additional resource: the number of qubits. We ask whether QViTs display similar scaling trends and whether accounting for this resource reveals different ways to allocate model capacity. In particular, can a larger qubit representation compensate for a smaller training parameter budget or less data? Parameter count and compute describe different aspects of training, so we examine the parameter trade-off and training runtime separately.

To address these questions, we numerically study QViTs across qubit numbers, transformer depths, and dataset sizes. A joint shifted power-law fit describes the observed validation losses and allows us to derive resource combinations that preserve a target loss. Within a regime identified by this fit, additional qubits can compensate for a reduction in classical encoder parameters. Earlier studies of quantum vision transformers and quantum deep hedging report competitive performance with fewer trainable parameters in the model components compared~\cite{Cherrat2024quantumvision,Cherrat2023DeepHedging}. Our scaling analysis gives a quantitative framework for investigating such parameter efficiency and suggests a way to investigate possible polynomial improvements in parameter efficiency: exchanging a larger qubit budget for a smaller parameter budget at comparable performance.

The same question also connects to classical neural-network theory. Under the unary encoding used here, qubit number is the representation width, and the quantum attention projections can be written as networks of sparsely connected pairwise rotations. Qubit number and parameter count thus jointly describe how trainable degrees of freedom are distributed across a structured network. Classical work on sparse matrix factorizations~\cite{Dao2019Butterfly} and the lottery ticket hypothesis~\cite{Frankle2019Lottery,Malach2020Lottery} suggests that appropriate structure can preserve predictive power with fewer parameters. This perspective may help explain the qubit--parameter trade-off and turn empirical QML scaling relations into more predictive theories.

\section{QViT Architecture and Experimental Setup}
In this section, we briefly introduce the QViT architecture considered in this work and describe the corresponding experimental setup.

\subsection{QViT Architecture}
The architecture studied here follows the unary-encoded quantum orthogonal attention formulation introduced in the original QViT proposal~\cite{Cherrat2024quantumvision}.
The main distinction between QViTs and classical vision transformers (ViTs) lies in the mechanism used to implement the attention computation. As shown in Fig.~\ref{fig:QViT_overview}, the classical attention module is replaced by a quantum attention module implemented by parameterized quantum circuits. In the orthogonal-transformer variant of the original QViT formulation~\cite{Cherrat2024quantumvision}, the bilinear attention score between patches $x_i$ and $x_j$ is written as
\[
A_{ij} = x_i^\top W x_j,
\]
where $W$ is a trainable orthogonal matrix. This formulation replaces the factorized form $W_Q W_K^\top$ with a single orthogonal transformation $W$. To better align with the standard attention mechanism, we use two separate quantum circuits for the query and key transformations, respectively.

On the other hand, we do not optimize the value projection matrix \(W_V\). Because our design already employs two separate trainable quantum circuits for the query and key transformations, making the value branch trainable as well would substantially increase the overall training overhead. For this reason, we keep the value branch fixed, so the encoding process shown in Fig.~\ref{fig:QViT_overview} is a fixed mapping circuit rather than a trainable one. In practice, this is equivalent to setting \(W_V = I\) and relying on the patch embedding layer to provide the representation used in the value branch. In a more general setting without such efficiency constraints, the encoding circuit could also be made trainable so that the value transformation is learned jointly with the query and key branches.

Following the original QViT construction~\cite{Cherrat2024quantumvision}, we use quantum orthogonal layers to parameterize the attention matrices. These layers are realized as quantum circuits built from two-qubit RBS gates arranged in a pyramid structure, as shown in Fig.~\ref{fig:QViT_overview}.

\subsection{Experimental Setup}

In principle, these quantum attention matrices may be executed on quantum hardware. However, doing so would be prohibitively time-consuming in our setting, especially since this work focuses on scaling laws and therefore requires a large number of training runs. For this reason, all training and evaluation reported here are performed by classical simulation.

A generic statevector simulation of an \(n\)-qubit circuit tracks \(2^n\) amplitudes. Repeated circuit evaluations and gradient calculations can therefore be costly across a large training sweep, even without constructing a dense unitary matrix. Our experiments use up to \(25\) qubits, and we exploit the circuit structure to reduce this simulation cost.

To overcome this difficulty, we exploit the special structure of the circuits adopted in QViT~\cite{Cherrat2024quantumvision}. The input is loaded using unary amplitude encoding, namely a basis of Hamming weight \(1\), where exactly one qubit is in state \(\ket{1}\) and all others are in state \(\ket{0}\). Moreover, the quantum orthogonal layers are built from RBS gates, which are Hamming-weight-preserving. As a result, if the input state lies in the Hamming-weight-\(1\) subspace, the entire computation remains inside the same subspace throughout the circuit.

We therefore restrict the simulation to the Hamming-weight-\(1\) subspace. The state has only \(n\) amplitudes in this basis, and the circuit action can be represented by an \(n\times n\) orthogonal matrix. This reduced representation enables the numerical scaling study. It also provides a useful classical description of the attention module: each RBS gate acts as a two-coordinate rotation, and the pyramid circuit specifies how these rotations are connected. We return to this structured-network interpretation in Sec.~\ref{sec:sparsity}.

Under the unary amplitude encoding used in this work, a \(d\)-dimensional input vector is encoded into the Hamming-weight-1 subspace of an \(n\)-qubit system. Because this subspace is \(n\)-dimensional, we choose \(n=d\). Under the unary-amplitude-encoding setup considered here, the number of qubits therefore coincides with the embedding dimension. Moreover, for the pyramid circuit architecture considered here, the corresponding \(n\)-qubit quantum circuit involves \(\frac{n(n-1)}{2}\) trainable parameters. 

In this work, we study the relationship between model performance and several important factors, including the number of qubits, dataset size, transformer layers, and a classical-simulation runtime proxy. To this end, we construct image-classification datasets from WebVision~\cite{Li2017WebVision}. The dataset sizes considered in our experiments are \(5\mathrm{k}\), \(10\mathrm{k}\), \(20\mathrm{k}\), and \(30\mathrm{k}\). The maximum number of training epochs is set to 200. In addition, we consider transformer architectures with 3, 6, and 12 layers, while the number of qubits is varied over \(\{5,10,15,20,25\}\). To control the computational cost of the experimental sweep, we use a single attention head throughout. The batch size is set to 32. All experiments are conducted on an NVIDIA A100 GPU.

\section{Scaling Laws for QViT}
In this section, we investigate how QViT performance depends on qubit number, training data, and model size. We first examine qubit number, dataset size, and transformer depth separately and then fit their joint effect on validation loss. We use this fit to identify how qubits can be exchanged for parameters or data at comparable performance, discuss a structured-sparsity interpretation, and examine training compute through a runtime proxy.

\subsection{Effect of the Number of Qubits}

We first study how the number of qubits affects model performance. As illustrated in Fig.~\ref{fig:scaling_factors}(a), increasing the number of qubits generally leads to lower validation loss, especially when moving from very small quantum representations to medium-sized ones. This observation indicates that enlarging the qubit-aligned embedding dimension can enhance the representational capacity of QViTs.

At the same time, the gain from increasing the number of qubits is not uniform across the entire range. In particular, increasing the number of qubits from 5 to 10 consistently yields a substantial improvement. However, beyond this regime, the benefit of adding more qubits becomes less stable and less pronounced, suggesting a gradual saturation of model capacity. These results suggest that the number of qubits should not be scaled blindly, but rather selected in a way that is commensurate with both task complexity and transformer layers.
\subsection{Effect of Dataset Size}

We next analyze the role of dataset size. Fig.~\ref{fig:scaling_factors}(b) shows a clear and stable trend: increasing the dataset size consistently reduces the validation loss across different qubit numbers and transformer layers.
\subsection{Effect of Transformer Layers}

\begin{figure*}[t]
    \centering
    \begin{minipage}{0.32\textwidth}
        \centering
        \includegraphics[width=\linewidth]{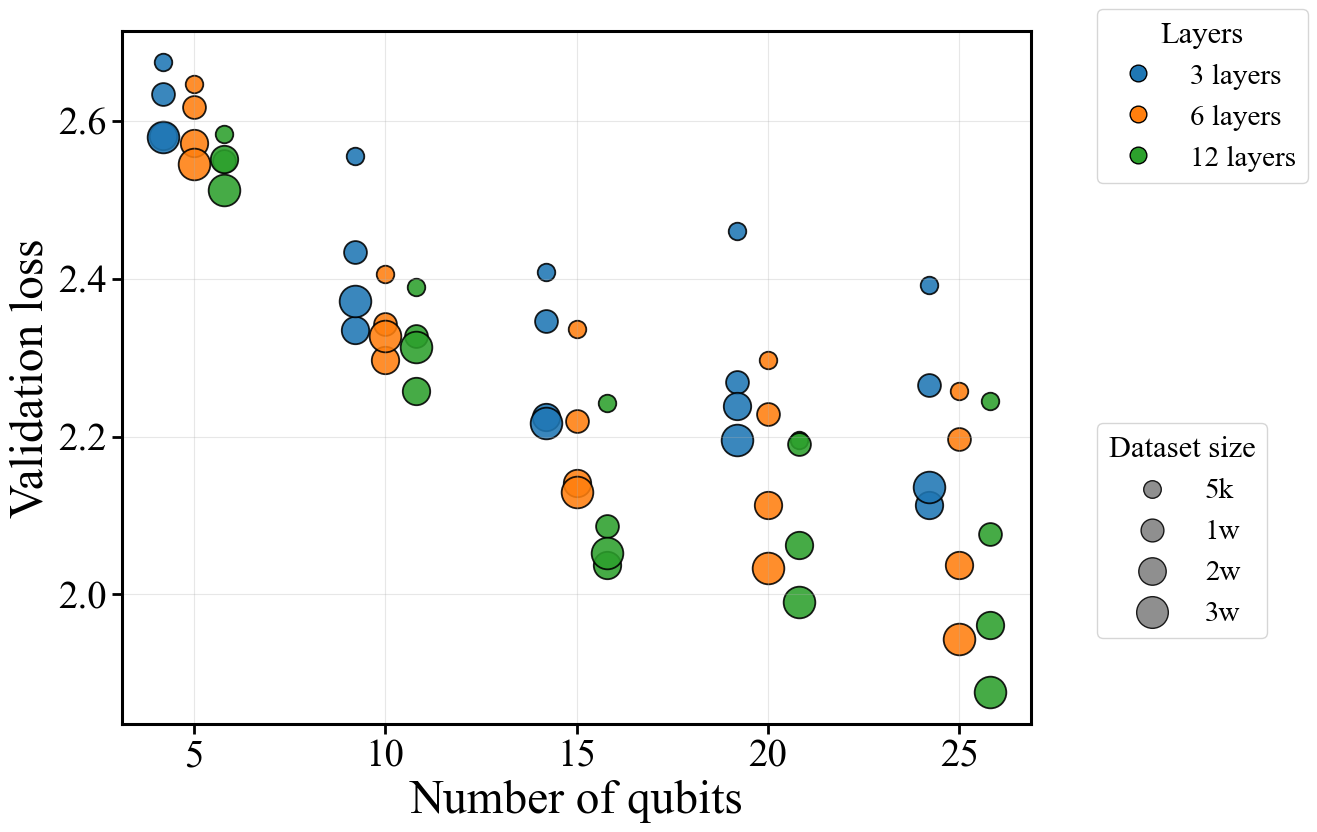}\\
        \textbf{(a)}
    \end{minipage}
    \hfill
    \begin{minipage}{0.32\textwidth}
        \centering
        \includegraphics[width=\linewidth]{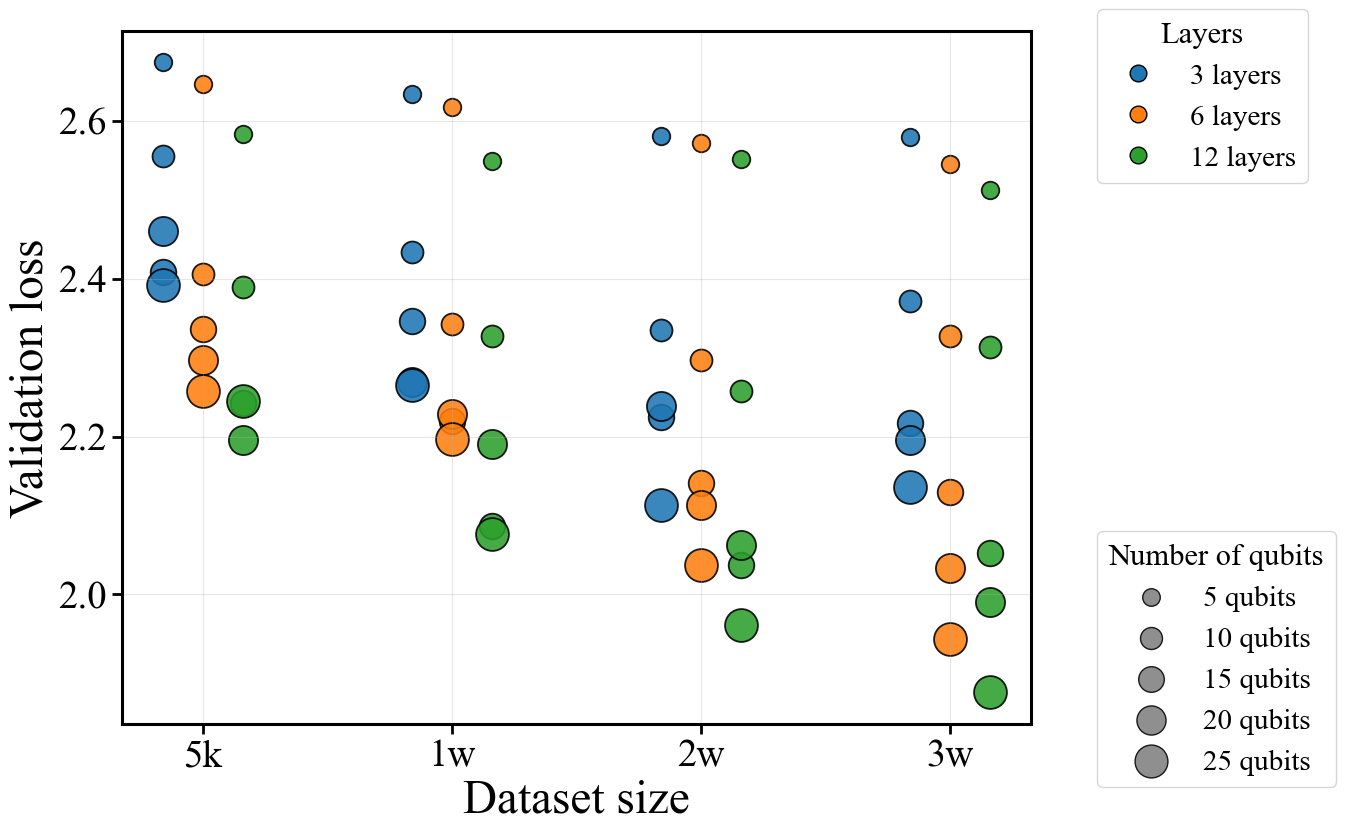}\\
        \textbf{(b)}
    \end{minipage}
    \hfill
    \begin{minipage}{0.32\textwidth}
        \centering
        \includegraphics[width=\linewidth]{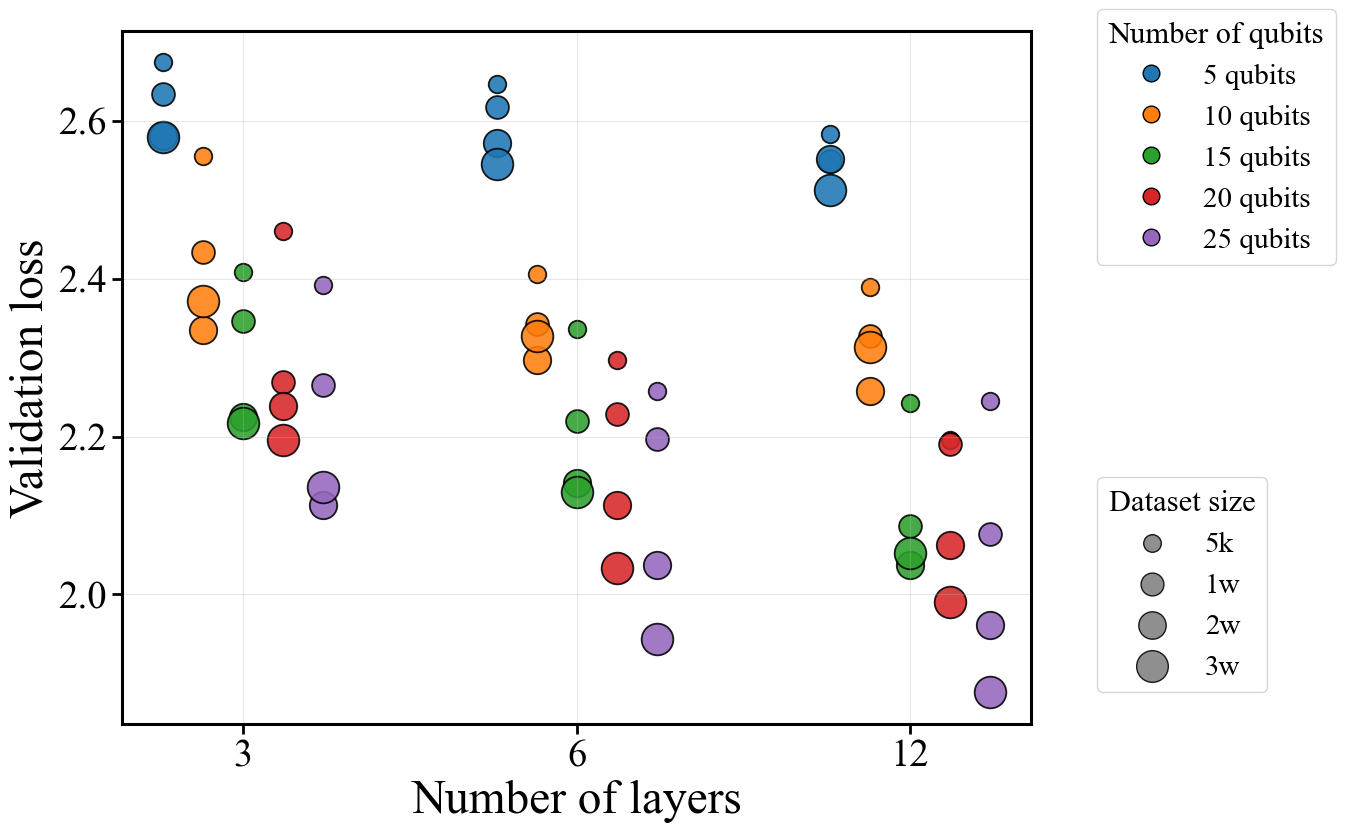}\\
        \textbf{(c)}
    \end{minipage}
    \caption{Validation loss as a function of the primary scaling factors considered in this work: (a) the number of qubits, (b) training dataset size, and (c) transformer depth.}
    \label{fig:scaling_factors}
\end{figure*}

We then investigate the influence of transformer layers on the empirical performance of QViTs. As shown in Fig.~\ref{fig:scaling_factors}(c), increasing the number of layers generally leads to lower validation loss across most experimental settings. However, although the number of layers is doubled at each step, the decrease in validation loss remains relatively modest. In comparison with the trends associated with qubit number and dataset size, 
increasing the number of transformer layers provides a clearly smaller benefit. 
This indicates that increasing transformer layers still improves performance, but the marginal gain is more limited than that of the other two factors in our experimental regime.

\subsection{Joint Effects of Transformer Layers, Dataset Size, and Qubit Number}
To further quantify the empirical trends observed above, we fit the final validation loss using a simple three-factor scaling model. Power-law models, sometimes with a fitted loss floor, describe diminishing returns with increasing model size, data, and compute in classical neural scaling studies~\cite{Hestness2017Scaling,Kaplan2020ScalingLaws,Zhai2022ScalingViT}. Motivated by this convention, we use an additive separable form in which each factor contributes a power-law correction to a shared loss floor:
\begin{equation}
\label{eq:joint_scaling}
L(q,\ell,D)=L_{\infty}+A_q q^{-\alpha_q}+A_\ell \ell^{-\alpha_\ell}+A_D D^{-\alpha_D},
\end{equation}
where \(q\) denotes the number of qubits, \(\ell\) the number of transformer layers, and \(D\) the dataset size. Here, \(L_{\infty}\) is a fitted loss-floor parameter, \(A_q,A_\ell,A_D>0\) are fitted amplitudes, and the positive exponents \(\alpha_q\), \(\alpha_\ell\), and \(\alpha_D\) characterize the scaling behavior with respect to the three factors.

The additive form provides a parsimonious first-order summary of the three controlled factors and permits their empirical contributions to be interpreted separately. It is not derived as a unique theoretical law for QViTs and does not model higher-order interactions among \(q\), \(\ell\), and \(D\). Accordingly, we use it to describe the finite experimental regime considered here rather than to assert a universal or asymptotic scaling form.

The fitted model provides a reasonably good in-sample approximation to the observed configurations after 200 epochs, with \(R^2=0.935\), RMSE \(=0.050\), and MAE \(=0.040\). This suggests that a simple shifted power-law form is able to capture the main variation of the validation loss at this stage across the considered model and data scales. Given the limited number of experimental configurations in our study, however, this fitted law should be regarded as an empirical summary of the observed regime rather than as a precise asymptotic scaling formula.

The fitted exponents are \(\alpha_q=0.496\), \(\alpha_\ell=0.457\), and \(\alpha_D=0.452\). Their positive values capture diminishing loss contributions as each resource increases, consistent with Fig.~\ref{fig:scaling_factors}. The shifted power-law form parallels classical neural scaling laws, while its explicit qubit dependence makes it possible to study how a larger quantum representation changes the allocation of parameters and data. The benefit of changing a resource depends on both its fitted amplitude and the operating scale. We therefore use the full fitted model, rather than the exponents alone, to quantify these trade-offs.

\subsection{Trade-off between Qubit Number and Classical Parameters}
\label{sec:parameter_tradeoff}

To quantify whether more qubits can compensate for fewer parameters, we rewrite the fitted loss in terms of the classical encoder parameter count. With embedding dimension \(E=q\) and MLP hidden dimension \(H=4E\), the classical parameters in \(\ell\) encoder blocks number
\begin{equation}
M_c=\ell f(q), \qquad f(q)=8q^2+9q.
\label{eq:classical_parameters}
\end{equation}
Here \(M_c\) counts the trainable classical parameters in the encoder blocks. The two query and key pyramid circuits additionally contribute \(M_q=\ell q(q-1)\) trainable angles, giving a total encoder-block count \(M_c+M_q=\ell(9q^2+8q)\). Patch embedding and the prediction head contribute further parameters. We first analyze \(M_c\), then explain how to include the circuit angles in the same construction.

Substituting \(\ell=M_c/f(q)\) into Eq.~\eqref{eq:joint_scaling} gives
\begin{equation}
\begin{split}
\mathcal{L}(q,M_c,D)={}&L_{\infty}+A_q q^{-\alpha_q}\\
&+A_\ell\left(\frac{M_c}{f(q)}\right)^{-\alpha_\ell}
+A_D D^{-\alpha_D}.
\end{split}
\label{eq:parameter_scaling}
\end{equation}
For a fixed dataset size and target loss \(L_t\), define \(\Delta=L_t-L_{\infty}-A_DD^{-\alpha_D}\). The resulting iso-loss relation is
\begin{equation}
M_c(q;L_t,D)=f(q)
\left[\frac{A_\ell}{\Delta-A_q q^{-\alpha_q}}\right]^{1/\alpha_\ell},
\label{eq:parameter_isoloss}
\end{equation}
provided \(\Delta>A_q q^{-\alpha_q}\). This expression retains the linear term in the parameter count, which is appreciable at the smallest qubit numbers. It describes a continuous interpolation of the fitted loss; realizable designs must also respect the discrete depths and qubit counts in the experimental sweep.

Along this fixed-loss curve, the local trade-off is
\begin{equation}
\frac{d\log M_c}{d\log q}
=\frac{qf'(q)}{f(q)}
-\frac{\alpha_q}{\alpha_\ell}
\frac{A_q q^{-\alpha_q}}{\Delta-A_q q^{-\alpha_q}}.
\label{eq:parameter_slope}
\end{equation}
A negative slope means that increasing the qubit count permits a smaller classical encoder parameter budget at the same predicted loss. In this regime, the gain from a wider representation compensates for the reduction in transformer depth needed to lower the parameter count. The budget \(M_c^\star(q)\) marks the point at which these two effects balance and the local slope is zero. Parameter savings are possible above this threshold, \(M_c>M_c^\star(q)\), where
\begin{equation}
M_c^\star(q)=f(q)\left[
\frac{\alpha_\ell A_\ell}{\alpha_q A_q}
\frac{qf'(q)}{f(q)}q^{\alpha_q}
\right]^{1/\alpha_\ell}.
\label{eq:exact_threshold}
\end{equation}
Keeping the dominant quadratic term in the parameter count, \(f(q)\simeq8q^2\), gives a simple expression for this threshold:
\begin{equation}
\begin{split}
M_c^\star(q)&\simeq Kq^{2+\alpha_q/\alpha_\ell},\\
K&=8\left(\frac{2\alpha_\ell A_\ell}{\alpha_q A_q}\right)^{1/\alpha_\ell}.
\end{split}
\label{eq:polynomial_threshold}
\end{equation}
The fitted exponents give \(2+\alpha_q/\alpha_\ell\simeq3.085\), with fitted normalization \(K\simeq1.83\). Thus, Eq.~\eqref{eq:polynomial_threshold} specifies the parameter budget above which increasing the qubit number can locally save parameters. For a chosen target loss, the amount that can be saved is determined by Eq.~\eqref{eq:parameter_isoloss} and depends on the starting configuration.

For the tested counts \(q\in\{5,10,15,20,25\}\) and \(\ell\in\{3,6,12\}\), the leading-order boundary places all configurations above the threshold except \((q,\ell)=(15,3),(20,3),(25,3),(25,6)\). The fitted model therefore predicts local parameter substitution in part of the explored regime, rather than throughout it. The same iso-loss construction can be applied to the total encoder-block count by replacing \(f(q)\) with \(9q^2+8q\); its leading-order boundary has the same exponent but a different prefactor. Full-model comparisons additionally require the embedding and prediction-head parameters.

These relations provide a quantitative expression of the parameter-efficiency mechanism motivating this work. A wider qubit representation can allow a shallower encoder with fewer classical parameters while retaining comparable predicted performance. This is consistent with earlier parameter-efficiency observations in quantum learning~\cite{Cherrat2024quantumvision,Cherrat2023DeepHedging} and motivates further study of possible parameter savings in quantum models. The power-law threshold identifies where this mechanism is available within the fitted model.

\subsection{Trade-off between Qubit Number and Dataset Size}

The qubit coordinate also makes it possible to ask whether a wider representation can compensate for less training data. At fixed depth \(\ell\) and target loss \(L_t\), define \(\Delta_D=L_t-L_{\infty}-A_\ell\ell^{-\alpha_\ell}\). Equation~\eqref{eq:joint_scaling} then gives
\begin{equation}
D(q;L_t,\ell)=
\left[\frac{A_D}{\Delta_D-A_q q^{-\alpha_q}}\right]^{1/\alpha_D},
\label{eq:data_isoloss}
\end{equation}
provided \(\Delta_D>A_q q^{-\alpha_q}\). Differentiating along this curve yields
\begin{equation}
\frac{dD}{dq}
=-\frac{\alpha_q A_q}{\alpha_D A_D}
q^{-\alpha_q-1}D^{\alpha_D+1}<0.
\label{eq:data_slope}
\end{equation}
Thus, within the fitted model, more qubits can partially offset a smaller dataset at fixed predicted loss and depth. Equation~\eqref{eq:data_isoloss} quantifies this exchange for a chosen target loss. Because increasing \(q\) at fixed depth also increases the parameter count, the data trade-off should be read together with the parameter trade-off in Sec.~\ref{sec:parameter_tradeoff}.

\subsection{Qubit Number, Parameter Budget, and Structured Sparsity}
\label{sec:sparsity}

From a classical neural-network viewpoint, the QViT attention projections are structured sparse networks of two-coordinate rotations. Each RBS gate mixes two coordinates of a \(q\)-dimensional representation and leaves the remaining coordinates unchanged. Their composition can produce a dense orthogonal transformation through a sparse computational graph. This connects the architecture to classical approaches that learn structured linear transformations as products of sparse factors~\cite{Dao2019Butterfly}.

In this description, qubit number sets the representation width, while the parameter budget controls the number and arrangement of trainable degrees of freedom across encoder blocks. For example, fixing \(M_c\) imposes \(\ell=M_c/(8q^2+9q)\), so a larger qubit representation uses fewer transformer blocks. The pair consisting of qubit number and parameter count can therefore help characterize the allocation and density of trainable connections relative to a specified dense reference network. The notion of sparsity here concerns the factorized attention architecture; the full QViT also contains dense components. A numerical sparsity measure would require fixing the reference architecture and specifying which connections are counted.

The parameter-substitution regime suggests that increasing representation width can preserve predictive performance even as the classical encoder parameter budget falls. Classical theories of sparse approximation, initialization, and optimization may help explain when this is possible. In particular, the lottery ticket hypothesis and related theoretical results~\cite{Frankle2019Lottery,Malach2020Lottery} motivate studying which structured connections carry the useful predictive capacity of QViTs. Such an analysis could provide a theoretical explanation for the empirical trade-off and its dependence on the task and architecture.

A related quantum perspective comes from symmetric pruning: for quantum neural networks used in ground-state preparation, exploiting symmetry and a smaller effective state space can reduce the circuit resources needed for favorable training~\cite{Wang2023SymmetricPruning}. This motivates examining how the restricted state space and connectivity of QViT attention affect parameter efficiency. Together with quantum neural tangent-kernel analyses of training dynamics~\cite{Liu2022QNTK,Liu2023WideQNN}, this offers a possible starting point for a theory of the observed trade-off; relating training dynamics to validation-loss scaling remains a further step.

\subsection{Effect of Classical-Simulation Runtime Proxy}

The above analyses compare different configurations after a common budget of 200 epochs. At fixed batch size, larger datasets require more optimizer steps per epoch, so the fitted dataset effect reflects this training protocol as well as the amount of data. Different architectures also incur different per-step costs in our reduced-space simulator. To provide a complementary view, we examine the training dynamics of QViTs using a classical-simulation runtime proxy.

\begin{figure}[t]
    \centering
    \includegraphics[width=\linewidth]{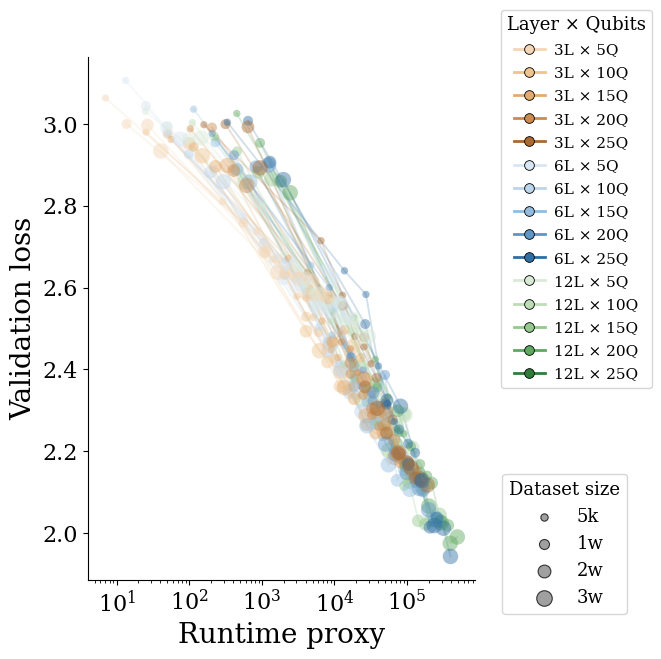}
    \caption{Validation loss as a function of the reduced-space classical-simulation wall-clock proxy.}
    \label{fig:runtime_proxy}
\end{figure}

We estimate the runtime proxy from the average wall-clock time per training step in the reduced-space implementation, using the same software and NVIDIA A100 environment across runs. It provides a complementary measure of the training cost associated with different parameter and qubit budgets. This proxy refers to the numerical implementation; execution on quantum hardware would require additional accounting for state preparation, circuit sampling, and gradient evaluation. Figure~\ref{fig:runtime_proxy} shows how validation loss evolves with this training budget.

Overall, the validation loss decreases as the classical-simulation runtime proxy increases, indicating that larger effective simulation budgets generally lead to better performance. At the same time, different configurations follow different trajectories. Larger models often require a larger simulated runtime budget before their benefits become visible, whereas smaller models may improve more quickly at the beginning but saturate earlier.

\section{Discussion and Conclusion}
QML promises new approaches to learning, but identifying where quantum resources improve predictive performance or resource efficiency remains a central theoretical challenge. We have approached this question through empirical scaling laws of QViTs, an architecture closely connected to the transformers underlying modern LLMs. The numerical results show that validation loss follows diminishing-return trends with qubit number, transformer depth, and training dataset size. A joint shifted power-law model captures these trends and extends the familiar classical scaling perspective by tracking qubit number explicitly.

The main implication is that qubit number can serve as a resource to trade against the training parameter budget. Within the regime identified in Sec.~\ref{sec:parameter_tradeoff}, a larger qubit representation can compensate for fewer classical encoder parameters at comparable predicted performance. The corresponding data relation describes how more qubits can also compensate for a smaller training set at fixed depth. These results motivate further study of whether such trade-offs can yield polynomial improvements in parameter efficiency. They are consistent with earlier benchmark results reporting competitive performance with fewer trainable parameters in the attention or model components compared~\cite{Cherrat2024quantumvision,Cherrat2023DeepHedging} and turn that observation into a quantitative resource-allocation question.

The structured-sparsity perspective offers a possible explanation. In a classical description, the attention projections form networks of pairwise rotations with a prescribed connectivity pattern. Qubit number determines their width, and the parameter budget determines how trainable transformations are allocated across the model. Together, these quantities provide a way to investigate the effective sparsity and structure of the learning architecture. A wider representation with an appropriate pattern of trainable interactions may retain useful features while requiring fewer classical encoder parameters.

This possibility resonates with the lottery ticket hypothesis: selected sparse subnetworks, with suitable initializations, can be trained to match the performance of their larger parent networks~\cite{Frankle2019Lottery}. Related theory establishes approximation guarantees for subnetworks contained in sufficiently overparameterized random ReLU networks under boundedness assumptions~\cite{Malach2020Lottery}. Together with work on structured sparse factorizations~\cite{Dao2019Butterfly}, these results suggest that classical neural-network theory may explain why qubit number can trade off against training parameters. Adapting these results to the connectivity and initialization of QViTs could provide a theoretical explanation of the observed resource exchanges and scaling exponents.

Future work can test this interpretation by comparing QViTs with dense and structured sparse classical transformers, varying circuit connectivity and initialization, and measuring parameter efficiency at fixed predictive performance. Larger scaling studies and validation of the predicted resource exchanges would clarify how broadly the fitted relations apply. Connecting these empirical laws to theories of approximation, sparsity, and optimization may help clarify when resource savings are possible and guide the design of more effective quantum learning models.

\begin{acknowledgments}
YQL, XJ, and JYL is supported in part by the University of Pittsburgh, School of Computing and Information, Department of Computer Science, Pitt Cyber, Pitt Momentum fund, PQI Community Collaboration Awards, John C. Mascaro Faculty Scholar in Sustainability, Switzerland NSF award 2000-1-243053, NSF award 2535915, 2610010, 2239175, DOE Genesis Program, Thinking Machines Lab and Cisco Research. This research used resources of the Oak Ridge Leadership Computing Facility, which is a DOE Office of Science User Facility supported under Contract DE-AC05-00OR22725.
\end{acknowledgments}

\section*{Data Availability}

The WebVision data used to construct the benchmark subsets are described in Ref.~\cite{Li2017WebVision}.

%


\begin{thebibliography}{18}%
\makeatletter
\providecommand \@ifxundefined [1]{%
 \@ifx{#1\undefined}
}%
\providecommand \@ifnum [1]{%
 \ifnum #1\expandafter \@firstoftwo
 \else \expandafter \@secondoftwo
 \fi
}%
\providecommand \@ifx [1]{%
 \ifx #1\expandafter \@firstoftwo
 \else \expandafter \@secondoftwo
 \fi
}%
\providecommand \natexlab [1]{#1}%
\providecommand \enquote  [1]{``#1''}%
\providecommand \bibnamefont  [1]{#1}%
\providecommand \bibfnamefont [1]{#1}%
\providecommand \citenamefont [1]{#1}%
\providecommand \href@noop [0]{\@secondoftwo}%
\providecommand \href [0]{\begingroup \@sanitize@url \@href}%
\providecommand \@href[1]{\@@startlink{#1}\@@href}%
\providecommand \@@href[1]{\endgroup#1\@@endlink}%
\providecommand \@sanitize@url [0]{\catcode `\\12\catcode `\$12\catcode
  `\&12\catcode `\#12\catcode `\^12\catcode `\_12\catcode `\%12\relax}%
\providecommand \@@startlink[1]{}%
\providecommand \@@endlink[0]{}%
\providecommand \url  [0]{\begingroup\@sanitize@url \@url }%
\providecommand \@url [1]{\endgroup\@href {#1}{\urlprefix }}%
\providecommand \urlprefix  [0]{URL }%
\providecommand \Eprint [0]{\href }%
\providecommand \doibase [0]{https://doi.org/}%
\providecommand \selectlanguage [0]{\@gobble}%
\providecommand \bibinfo  [0]{\@secondoftwo}%
\providecommand \bibfield  [0]{\@secondoftwo}%
\providecommand \translation [1]{[#1]}%
\providecommand \BibitemOpen [0]{}%
\providecommand \bibitemStop [0]{}%
\providecommand \bibitemNoStop [0]{.\EOS\space}%
\providecommand \EOS [0]{\spacefactor3000\relax}%
\providecommand \BibitemShut  [1]{\csname bibitem#1\endcsname}%
\let\auto@bib@innerbib\@empty
\bibitem[{Liu et~al.(2021)}]{Liu2021RigorousSpeedup}
Y.~Liu, S.~Arunachalam, and K.~Temme,
\href{https://doi.org/10.1038/s41567-021-01287-z}{A rigorous and robust quantum speed-up in supervised machine learning},
Nat. Phys. \textbf{17}, 1013--1017 (2021).
\bibitem[{Huang et~al.(2021)}]{Huang2021PowerData}
H.-Y.~Huang, M.~Broughton, M.~Mohseni, R.~Babbush, S.~Boixo, H.~Neven, and J.~R.~McClean,
\href{https://doi.org/10.1038/s41467-021-22539-9}{Power of data in quantum machine learning},
Nat. Commun. \textbf{12}, 2631 (2021).
\bibitem[{Liu et~al.(2022)}]{Liu2022QNTK}
J.~Liu, F.~Tacchino, J.~R.~Glick, L.~Jiang, and A.~Mezzacapo,
\href{https://doi.org/10.1103/PRXQuantum.3.030323}{Representation learning via quantum neural tangent kernels},
PRX Quantum \textbf{3}, 030323 (2022).
\bibitem[{Liu et~al.(2023)}]{Liu2023WideQNN}
J.~Liu, K.~Najafi, K.~Sharma, F.~Tacchino, L.~Jiang, and A.~Mezzacapo,
\href{https://doi.org/10.1103/PhysRevLett.130.150601}{Analytic theory for the dynamics of wide quantum neural networks},
Phys. Rev. Lett. \textbf{130}, 150601 (2023).
\bibitem[{Kaplan et~al.(2020)}]{Kaplan2020ScalingLaws}
J.~Kaplan, S.~McCandlish, T.~Henighan, T.~B.~Brown, B.~Chess, R.~Child, S.~Gray, A.~Radford, J.~Wu, and D.~Amodei,
\href{https://arxiv.org/abs/2001.08361}{Scaling laws for neural language models},
arXiv:2001.08361 (2020).
\bibitem[{Cherrat et~al.(2024)}]{Cherrat2024quantumvision}
E.~A.~Cherrat, I.~Kerenidis, N.~Mathur, J.~Landman, M.~Strahm, and Y.~Y.~Li,
\href{https://doi.org/10.22331/q-2024-02-22-1265}{Quantum vision transformers},
Quantum \textbf{8}, 1265 (2024).
\bibitem[{Liu et~al.(2024)}]{Liu2024LargeScaleQML}
J.~Liu, M.~Liu, J.-P.~Liu, Z.~Ye, Y.~Wang, Y.~Alexeev, J.~Eisert, and L.~Jiang,
\href{https://doi.org/10.1038/s41467-023-43957-x}{Towards provably efficient quantum algorithms for large-scale machine-learning models},
Nat. Commun. \textbf{15}, 434 (2024).
\bibitem[{Boucher et~al.(2026)}]{Boucher2025qsaBiomed}
T.~Boucher, J.~Whittle, and E.~B.~Mazomenos,
\href{https://doi.org/10.1007/978-3-032-13961-0_12}{From $O(n^2)$ to $O(n)$ parameters: Quantum self-attention in vision transformers for biomedical image classification},
in \emph{Efficient Medical Artificial Intelligence}, LNCS \textbf{16318} (Springer, Cham, 2026), pp. 112--122.
\bibitem[{Comajoan Cara et~al.(2024)}]{Cara2024qvitQuarkGluon}
M.~Comajoan Cara, G.~R.~Dahale, Z.~Dong, R.~T.~Forestano, S.~Gleyzer, D.~Justice, K.~Kong, T.~Magorsch, K.~T.~Matchev, K.~Matcheva, and E.~B.~Unlu,
\href{https://doi.org/10.3390/axioms13050323}{Quantum vision transformers for quark-gluon classification},
Axioms \textbf{13}, 323 (2024).
\bibitem[{Unlu et~al.(2024)}]{Unlu2024hybridQViT_HEP}
E.~B.~Unlu, M.~Comajoan Cara, G.~R.~Dahale, Z.~Dong, R.~T.~Forestano, S.~Gleyzer, D.~Justice, K.~Kong, T.~Magorsch, K.~T.~Matchev, and K.~Matcheva,
\href{https://doi.org/10.3390/axioms13030187}{Hybrid quantum vision transformers for event classification in high energy physics},
Axioms \textbf{13}, 187 (2024).
\bibitem[{Hestness et~al.(2017)}]{Hestness2017Scaling}
J.~Hestness, S.~Narang, N.~Ardalani, G.~Diamos, H.~Jun, H.~Kianinejad, M.~M.~A.~Patwary, Y.~Yang, and Y.~Zhou,
\href{https://arxiv.org/abs/1712.00409}{Deep learning scaling is predictable, empirically},
arXiv:1712.00409 (2017).
\bibitem[{Zhai et~al.(2022)}]{Zhai2022ScalingViT}
X.~Zhai, A.~Kolesnikov, N.~Houlsby, and L.~Beyer,
\href{https://openaccess.thecvf.com/content/CVPR2022/html/Zhai_Scaling_Vision_Transformers_CVPR_2022_paper.html}{Scaling vision transformers},
in \emph{Proc. IEEE/CVF CVPR} (2022), pp. 12104--12113.
\bibitem[{Cherrat et~al.(2023)}]{Cherrat2023DeepHedging}
E.~A.~Cherrat, S.~Raj, I.~Kerenidis, A.~Shekhar, B.~Wood, J.~Dee, S.~Chakrabarti, R.~Chen, D.~Herman, S.~Hu, P.~Minssen, R.~Shaydulin, Y.~Sun, R.~Yalovetzky, and M.~Pistoia,
\href{https://doi.org/10.22331/q-2023-11-29-1191}{Quantum deep hedging},
Quantum \textbf{7}, 1191 (2023).
\bibitem[{Dao et~al.(2019)}]{Dao2019Butterfly}
T.~Dao, A.~Gu, M.~Eichhorn, A.~Rudra, and C.~R\'{e},
\href{https://proceedings.mlr.press/v97/dao19a.html}{Learning fast algorithms for linear transforms using butterfly factorizations},
in \emph{Proc. ICML}, PMLR \textbf{97}, 1517--1527 (2019).
\bibitem[{Frankle and Carbin(2019)}]{Frankle2019Lottery}
J.~Frankle and M.~Carbin,
\href{https://arxiv.org/abs/1803.03635}{The lottery ticket hypothesis: Finding sparse, trainable neural networks},
in \emph{ICLR} (2019).
\bibitem[{Malach et~al.(2020)}]{Malach2020Lottery}
E.~Malach, G.~Yehudai, S.~Shalev-Shwartz, and O.~Shamir,
\href{https://proceedings.mlr.press/v119/malach20a.html}{Proving the lottery ticket hypothesis: Pruning is all you need},
in \emph{Proc. ICML}, PMLR \textbf{119}, 6682--6691 (2020).
\bibitem[{Li et~al.(2017)}]{Li2017WebVision}
W.~Li, L.~Wang, W.~Li, E.~Agustsson, and L.~Van~Gool,
\href{https://arxiv.org/abs/1708.02862}{WebVision database: Visual learning and understanding from web data},
arXiv:1708.02862 (2017).
\bibitem[{Wang et~al.(2023)}]{Wang2023SymmetricPruning}
X.~Wang, J.~Liu, T.~Liu, Y.~Luo, Y.~Du, and D.~Tao,
\href{https://openreview.net/forum?id=K96AogLDT2K}{Symmetric pruning in quantum neural networks},
in \emph{ICLR} (2023).
\end{thebibliography}
\end{document}